\documentclass[11pt]{elsarticle}
\usepackage{amsfonts}
\usepackage{url}

\newtheorem{algorisme}{Algorithm}{\bfseries}{\itshape}
{\bfseries}{\itshape}
\newtheorem{lemma}{Lemma}{\bfseries}{\itshape}
\newtheorem{definition}{Definition}{\bfseries}{\itshape}
{\bfseries}{\itshape}
{\bfseries}{\itshape}
{\bfseries}{\itshape}

\begin{document}
\begin{frontmatter}
\title{Utility-Preserving Differentially Private Data Releases Via 
Individual Ranking Microaggregation}

\author{David S\'anchez}
\author{Josep Domingo-Ferrer}
\author{Sergio Mart\'{\i}nez}
\author{and Jordi Soria-Comas}
\address{UNESCO Chair in Data Privacy, Department of Computer Engineering 
and Mathematics, Universitat Rovira i Virgili, Av. Pa\"{\i}sos Catalans 26, E-43007 Tarragona, Catalonia\\
E-mail \{david.sanchez,josep.domingo,sergio.martinezl,jordi.soria\}@urv.cat}

\begin{abstract}
Being able to release and exploit open data gathered in
information systems is crucial for
researchers,
enterprises and the overall society. Yet, these data
must be anonymized before release to protect
the privacy of the subjects to whom the records relate.
Differential privacy is a privacy model for anonymization
that offers  
more robust privacy guarantees than previous
models, such as $k$-anonymity and its extensions.
However, it is often disregarded that 
the utility of differentially private outputs
is quite limited, either because of the amount of noise 
that needs to be added to obtain them or because 
utility is only preserved for a restricted type and/or a 
limited number of queries.
On the contrary, $k$-anonymity-like data releases make no assumptions 
on the uses of the protected data and, thus, do 
not restrict the number and type of doable analyses.
Recently, some authors have proposed mechanisms to offer general-purpose 
differentially
private data releases. 
This paper extends such works with a specific
focus on the preservation of the utility of the protected data.
Our proposal builds on microaggregation-based anonymization,
which is more flexible and utility-preserving than 
alternative anonymization methods used in the literature,
in order to reduce the amount of noise needed to satisfy differential 
privacy. In this way, we improve 
the utility of differentially private data releases. 
Moreover, the noise reduction we achieve does not depend on the 
size of the data set, but just on the number of attributes to be protected,
which is a more desirable behavior for large data sets.
The utility benefits brought by our proposal are empirically evaluated
and compared with related works for several data sets and metrics.
\\
 
\end{abstract}

\end{frontmatter}

\section{Introduction}

Releasing and exploiting open data is crucial to boost progress of 
knowledge, economy and society.
Indeed, the availability of such data facilitates research and 
allows
better marketing, better planning and better social services. 
However, data publication often faces privacy threats due to the
confidentiality of the information that is released for secondary use.
To tackle this problem, a plethora of methods
aimed at data anonymization have been proposed within the field 
of statistical disclosure control~\cite{Hundepool}.
Such methods distort input data in different ways ({\em e.g.} noise 
addition, removal, sampling, data generalization, etc.) 
so that the probability of
re-identifying individuals and, thus, disclosing their confidential information
is brought below a tolerable threshold. 
Even though those methods have been shown to improve
privacy protection while preserving 
a reasonable level of analytical utility
(the main motivation of data publishing), 
they offer no formal privacy guarantees. 

In contrast, privacy models proposed in recent years within the computer
science community~\cite{Drechsler,Danezis} seek to attain a predefined notion 
of privacy, thus offering {\em a priori} privacy guarantees. These guarantees
are interesting because they ensure a minimum level of privacy regardless
of the type of transformation performed on input data. Among such models,
$k$-anonymity and the more recent $\varepsilon$-differential privacy have
received a lot of attention. 

$k$-Anonym\-ity~\cite{SamaratiSweeney98,Samarati01} seeks 
to make each record in the input data set indistinguishable from, at least,
$k-1$ other records, so that the probability of re-identification of individuals
is, at most, $1/k$. 
Different anonymization methods have been proposed to achieve that goal, 
such as removal of outlying records, generalization
of values to a common
abstraction~\cite{Samarati01,Sweeney02,Aggarwal05,Goldberger10}
or multivariate microaggregation~\cite{Domi02,Domi05}. 
The latter method
partitions a data set into groups at least $k$ similar records and replaces
the records in each group by a prototypical record
({\em e.g.} the centroid record, that is, the average record).
Whatever the computational procedure, $k$-anonym\-ity 
focuses on masking quasi-identifier attributes; these are 
attributes ({\em e.g.}, Age, Gender, Zipcode and Race)
that are assumed to enable 
re-identifying the respondent of a record because they are linkable
to analogous attributes available in external identified data sources
(like electoral rolls, phone books, etc.). $k$-Anonym\-ity
does not mask confidential attributes ({\em e.g.}, salary,
health condition, political preferences, etc.) 
unless they are also quasi-identifiers. While 
$k$-anonymity has been shown to provide reasonably useful anonymized results,
especially for small $k$, it
is also vulnerable to attacks based on the possible lack of diversity
of the non-anonymized confidential attributes or on additional background
knowledge available to the attacker~\cite{Mach06,Wong06,Li07,Domipsai08}.

Unlike $k$-anonymity, the more recent 
$\varepsilon$-differential privacy~\cite{Dwork06} model does
not make any assumptions on which attributes are quasi-identifiers,
that is, on the background knowledge
available to potential attackers seeking to re-identify the respondent
of a record.
$\varepsilon$-Differential privacy guarantees that the anonymized 
output is insensitive (up to a factor dependent on $\varepsilon$) to 
the modification, deletion or addition of any single input record in 
the original data set. 
In this way, the privacy of any individual is 
not compromised by the publication of the anonymized output,
which is a much more robust guarantee than the one offered by $k$-anonymity.
Differential privacy is attained by adding
an amount of noise to the 
original outputs  
that is proportional to the \emph{sensitivity}
of such outputs to modifications of particular individual records
 in the original data set.
This sensitivity does not depend on the specific values of 
attributes in specific records, but on the domains 
of those attributes.
Basing the sensitivity and hence the added noise
on attribute domains rather than attribute values 
satisfies the privacy guarantee but may yield severely
distorted anonymized outputs, whose utility is very limited.
Because of this, $\varepsilon$-differential privacy was 
originally proposed for the 
\textit{interactive} scenario, in which the outputs
are the answers to interactive queries rather 
than the data set itself. When applying $\varepsilon$-differential
privacy to this scenario, 
the anonymizer returns noise-added answers to interactive queries.
In this way, the accuracy/utility of the response to a query depends on 
the sensitivity of the query, which is usually less than the sensitivity of the 
data set attributes. However, the interactive setting
of $\varepsilon$-differential privacy limits the number and type of queries 
that can be performed.
The proposed extensions 
of $\varepsilon$-differential privacy 
to the \textit{non-interactive} setting (generation of entire anonymized data sets) overcome the limitation
on the number of queries, but not on the type of queries
for which some utility is guaranteed (see Section~\ref{backdif} below).

\subsection{Contribution and plan of this paper}

In previous works~\cite{TrustCom,VLDB}, we showed that the noise required to
fulfill differential privacy in the non-interactive setting can be reduced 
by using a special
type of microaggregation-based $k$-anonymity on the input data set. 
The rationale is that the microaggregation
performed to achieve $k$-anonymity helps reducing the sensitivity 
of the input versus modifications of 
individual records.
As a result, data utility preservation can be improved 
(in terms of less data distortion) without renouncing the strong privacy
guarantee of differential privacy.
With such a mechanism, the sensitivity reduction depends on the number of
$k$-anonymized groups to be released; in turn, this number is a function of
the value of $k$ and the cardinality of the data set.
The larger the group size $k$, the less sensitive
are the group centroids resulting from the microaggregation; on the other hand, the smaller
the data set, the smaller the number of different group centroids in
the microaggregated data set. Thus, as the group size increases
or the data set size decreases, the sensitivity decreases and 
less noise needs to be added to reach differential privacy.
Hence, the resulting differentially private 
data have higher utility. 
In the two abovementioned works, we empirically showed that the noise 
reduction more than compensates the information loss introduced
by microaggregation.  

In line with~\cite{TrustCom,VLDB}, in this paper we investigate
other transformations of the original data 
aimed at reducing their sensitivity.
  The proposal in this paper is based on individual ranking microaggregation, a kind of microaggregation
  that is more flexible and utility-preserving than the one used in~\cite{TrustCom,VLDB}.
  In contrast to the previous works, 
the reduction of sensitivity achieved by the method presented in this paper 
does not depend on the size of the 
data set, but just on the number of attributes to be protected. 
This is more
desirable for large data sets or in scenarios in which only the confidential
attributes should be protected. 
In fact, experiments carried on two reference data sets show a significant
improvement of data utility (in terms of relative error and preservation of 
attribute distributions) with respect to the previous work.
Moreover, the microaggregation mechanism
used in this paper is simpler and more scalable, 
which facilitates implementation and practical
deployment.

The rest of this paper is organized as follows. 
Section~\ref{related} reviews background on
microaggregation,
$\varepsilon$-differential privacy and
$\varepsilon$-differentially private data publishing.
Section~\ref{dp_indiv} proposes the new method to generate 
$\varepsilon$-differentially private data sets that uses
a special type of microaggregation to reduce the amount of required noise. 
Implementation details are given for data sets
with numerical and categorical attributes.
Section~\ref{evaluation} reports an empirical comparison 
of the proposed method and previous proposals, based on two 
reference data sets.
The final section gathers some conclusions.

\section{Background and related work}
\label{related}

\subsection{Background on microaggregation}
\label{micro}

Microaggregation~\cite{Domi02,Defays93} is a family of anonymization
algorithms that works in two stages:
\begin{itemize}
\item First, the data set is clustered
in such a way that: i) each cluster contains at least $k$ elements; 
ii) elements within a cluster are as similar as possible.
\item Second, each element within each cluster is replaced by 
a representative of the cluster, typically the centroid value/tuple.
\end{itemize}

Depending on whether they deal with one or several attributes at a time,
microaggregation methods can be classified into univariate and multivariate:
\begin{itemize}
\item Univariate methods deal with multi-attribute data sets by microaggregating
one attribute at a time. Input records are sorted by the first attribute, then
groups of successive $k$ values of the first attribute 
are created and all values
within that group are replaced by the group representative ({\em e.g.}
centroid). 
The same procedure is repeated for the rest of attributes.
Notice that all attribute values of each record are moved together
when sorting records by a particular attribute; hence, the relation between
the attribute values within each record is preserved.
This approach is known as 
{\em individual ranking}~\cite{Defays93,Defays98} and,
since it microaggregates one attribute at a time, 
its output is not $k$-anonymous
at the record level.
Individual ranking just reduces the variability of attributes, 
thereby providing
some anonymization.
In~\cite{Domi01} it was shown that
individual ranking causes low information loss and, thus, its
output better preserves analytical utility. However,
the disclosure risk in the anonymized output remains 
unacceptably high~\cite{Domi02a}.
\item To deal with several attributes at a time,
the trivial option is to map multi-attribute data sets to univariate
data by projecting the former onto a single axis 
({\em e.g.} using the sum of z-scores or the first principal component, 
see~\cite{Defays98}) and then
use univariate microaggregation on the univariate data.
Another option avoiding the information loss due to 
single-axis projection is to 
use {\em multivariate microaggregation} able to deal 
with unprojected multi-attribute data~\cite{Domi02,Sande01}. 
If we define optimal microaggregation as finding a partition 
in groups of size at least $k$ such that within-groups homogeneity
is maximum, it turns out that, while optimal univariate
microaggregation can be solved in polynomial time~\cite{Hansen03},
unfortunately optimal multivariate microaggregation is NP-hard~\cite{Oganian01}.
This justifies the use of heuristic methods for multivariate
microaggregation, such as the MDAV algorithm~\cite{Domi05},
which has been extensively used to enforce $k$-anonymity 
at the record level~\cite{Domi05,Domingo10,Martinez12a}.
In any case, multivariate microaggregation leads to higher information loss than
individual ranking~\cite{Domi01}.
\end{itemize}

\subsection{Background on differential privacy}
\label{backdif}

Differential privacy was originally proposed by~\cite{Dwork06} as a privacy
model in the interactive setting, that is, to protect the outcomes
of queries to a database. The assumption is that an anonymization mechanism
sits between the user submitting queries and the database answering them.

\begin{definition}[$\varepsilon$-Differential privacy]
\label{def:diff_priv}
A randomized function $\kappa$
gives $\varepsilon$-differential
privacy if, for all data sets $X_{1}$, $X_{2}$ such that one can
be obtained from the other by modifying a single record,
and all $S\subset Range(\kappa)$, it holds
\begin{equation}
\label{eqprob}
\Pr(\kappa(X_{1})\in S)\le\exp(\varepsilon)\times \Pr(\kappa(X_{2})\in S).
\end{equation}
\end{definition}

The computational mechanism to attain $\varepsilon$-differen\-tial privacy
is often called $\varepsilon$-differentially private {\em sanitizer}.
A usual sanitization approach  
is noise addition: first, the real value $f(X)$ of the response to 
a certain user query $f$ is
computed, and then a random noise, say $Y(X)$, is added to mask $f(X)$, 
that is, a randomized
response $\kappa(X)= f(X) + Y(X)$ is returned. To generate
$Y(X)$, a common choice is to use a Laplace
distribution with zero mean and
$\Delta(f)/\varepsilon$ scale parameter, where:
\begin{itemize}
\item $\varepsilon$ is the differential privacy parameter; 
\item $\Delta(f)$ is the $L_1$-sensitivity of $f$, that is,
the maximum variation of the query function between neighbor
data sets, {\em i.e.}, sets differing in at most one record.
\end{itemize}

Specifically, the density function of the Laplace noise is 
\[ p(x) = \frac{\varepsilon}{2\Delta(f)} e^{-|x| \varepsilon/\Delta(f)}. \]
Notice that, for fixed $\varepsilon$, the higher the sensitivity $\Delta(f)$
of the query function $f$, the more Laplace noise is added: 
indeed, satisfying the $\varepsilon$-differential privacy 
definition (Definition~\ref{def:diff_priv}) requires more noise
when the query function $f$ can vary strongly between neighbor data sets.
Also, for fixed $\Delta(f)$, the smaller $\varepsilon$,
the more Laplace noise is added: when $\varepsilon$ is very
small, Definition~\ref{def:diff_priv} almost requires that 
the probabilities on both sides of Equation (\ref{eqprob}) be equal,
which requires the randomized function $\kappa(\cdot) = f(\cdot) + Y(\cdot)$ 
to yield very similar results for all pairs of neighbor data sets; adding
a lot of noise is a way to achieve this.

Differential privacy was also proposed for the non-interactive setting 
in~\cite{Blum,Dwor09,Hardt2010,Chen11}. Even though a 
non-interactive data release
can be used to answer an arbitrarily large number of queries, in all
these proposals, this is obtained at the cost of 
offering utility guarantees only for a 
restricted class of queries~\cite{Blum}, typically count queries.

\subsection{Related work on differentially private data publishing}
\label{relwork}

In contrast to the general-purpose data publication offered 
by $k$-anonymity, which makes no assumptions on the uses of published data
and does not limit the type and number of analyses that can be performed,
$\varepsilon$-differential privacy severely limits data uses.
Indeed, in the interactive scenario, $\varepsilon$-differential
privacy allows only a limited and pre-selected number of queries
of a certain type to be answered; 
in the extensions to the non-interactive scenario, 
any number of queries can be answered, but utility guarantees
are only offered for a restricted class of queries. 
We next review the literature on non-interactive 
$\varepsilon$-differential privacy, which is the focus
of this paper.

The usual approach to releasing 
differentially private data sets 
is based on histogram queries~\cite{Xiao2010b,Xu2012}, that is, 
on approximating the data distribution by partitioning
the data domain and counting the number of records in each partition set.
To prevent
the counts from leaking too much information 
they are computed in a differentially private manner.
Apart from the counts, partitioning can also reveal information. 
One way to prevent 
partitioning from leaking information consists in using 
a predefined partition that is independent
of the actual data under consideration ({\em e.g.} by 
using a grid~\cite{Mach08}). 

   The accuracy of the approximation obtained via histogram 
queries depends on the size
   of the histogram bins (the greater they 
are, the more imprecise is the attribute value)
   as well as on the number of records contained in them (the more records, the less relative
   error). 
For data sets with sparsely populated regions, using a predefined 
partition may be problematic. Several strategies have been proposed to improve the accuracy of differentially private count (histogram) queries, which
we next review. In~\cite{Hay2010} consistency constraints
between a set of queries are exploited to increase accuracy. 
In~\cite{Xiao2010} a wavelet transform is performed on the data and noise is added in the
frequency domain. In~\cite{Xu2012,Li13} the histogram bins are adjusted to the actual data. In~\cite{Cormode2012}, the authors consider
differential privacy of attributes
whose domain is ordered and has moderate to large cardinality
({\em e.g.} numerical attributes); the attribute domain
is represented as a tree, which is decomposed in order to
increase the accuracy of answers to count queries (multi-dimensional
range queries).
In \cite{Mohammed11}, the authors generalize similar records by using coarser categories
for the classification attributes; this results in higher counts
of records in the histogram bins, which are much larger than the noise that
needs to be added to reach differential privacy.
 For data sets with a significant number of attributes, 
attaining differential privacy
 while at the same time preserving the accuracy of the attribute values 
(by keeping 
the histogram bins
 small enough) becomes a complex task. Observe that, given a 
number of bins per attribute,
 the total number of bins grows exponentially with the number of attributes. 
Thus, in order to avoid obtaining 
 too many sparsely populated bins, the number of bins per attribute must 
 be significantly reduced (with the subsequent accuracy loss).
 An interesting approach to deal with multidimensional data is proposed in~\cite{Mir,Zhang}.
 The goal of these papers is to compute differentially private histograms independently
 for each attribute (or jointly for a small number of attributes) and then try to generate
 a joint histogram for all attributes from the partial histograms. 
This was done for a data set of commuting patterns in~\cite{Mir}
 and for an arbitrary data set in~\cite{Zhang}. In particular,~\cite{Zhang} 
first tried to build a
 dependency hierarchy between attributes. Intuitively, when two attributes 
are independent,
 their joint histogram can be reconstructed from the histograms of each of the attributes; thus,
 the dependency hierarchy helps determining which marginal or low-dimension histograms 
 are more interesting to approximate the joint histogram. The approaches in~\cite{Mir,Zhang}
 can be seen as complementary to our proposal, which can be used as an
alternative
 for computing the histograms. This does not mean that our proposal is not competitive
 against~\cite{Mir,Zhang} in terms of data utility. The data utility depends highly on the actual
 data that we have. The most favorable cases for~\cite{Zhang} are data sets with low dependency
 between attributes (for instance, when attributes are completely independent, computing the marginal histograms is enough)
 or at least with attributes that depend on a small number of other attributes (for example, for an attribute
 that depends only on another attribute, the joint histogram for these two attributes is enough).
 On the contrary, if an attribute has a sizeable and similar 
level of dependency on all the other attributes, there is no
 advantage in using~\cite{Zhang}.
 
Our work differs from all previous ones
in that it is not limited to histogram queries and it allows
dealing with any type of attributes (ordered or unordered).

\subsection{Related work on microaggregation-based differentially-private data publishing}
\label{relwork2}

In ~\cite{VLDB} we presented an approach that combines $k$-anonymity and
$\varepsilon$-differential privacy in order to reap the 
best of both models: namely, the 
reasonably low information loss incurred by $k$-anonymity
and its lack of assumptions on data uses,
as well as the robust privacy guarantees offered by $\varepsilon$-differential
privacy.
In that work, we first defined the notion of insensitive microaggregation,
which is a multivariate microaggregation procedure that partitions 
data in groups of $k$ records with a criterion that 
  is relatively insensitive to changes in the data set.
To do so, insensitive microaggregation 
defines a total order for the joint domains of all the attributes of
the input data set $X$.
Insensitive microaggregation ensures that, for every pair of 
data sets $X$ and $X'$ differing
in a single record, the resulting clusters will
differ at most in a single record. Hence, the centroids used to
replace records of each cluster will have low sensitivity to changes
of one input record. Specifically, when centroids are computed as 
the arithmetic average 
of the elements of the cluster, the sensitivity is as low as $\Delta(X)/k$,
where $\Delta(X)$ is the distance between the most distant records of the joint domains 
of the input data and $k$ is the 
size of the clusters. This sensitivity is much lower and
restrained than the one offered by standard microaggregation algorithms, 
such as MDAV~\cite{Domi05}, whose output is highly dependent on the input data ({\em i.e.}  
modification of a single record may lead to completely different clusters). 
Note also that the sensitivity of individual input records 
({\em i.e.} without microaggregation)
is $\Delta(X)$.
The downside of insensitive microaggregation is
that it yields worse within-cluster
homogeneity than standard microaggregation 
and, hence, higher information loss.

As a result of insensitive microaggregation with cluster cardinality $k$,
input data are $k$-anonymized in such a way
that all attributes
are considered as being quasi-identifiers. To obtain a differentially private
output, an amount of noise needs to be added to cluster centroids 
that depends on their sensitivity.
Centroids provided by insensitive microaggregation 
have a low sensitivity and thus require little noise, which in
turn means incurring low information loss.
In data publishing, $n/k$ centroids are released, 
each one computed on a cluster of cardinality $k$ and having sensitivity 
$\Delta(X)/k$ (see discussion above). 
Hence, the sensitivity of the whole data set to be released is
$n/k\times \Delta(X)/k$. Thus, for numerical data sets, 
Laplace noise with scale parameter $(n/k\times \Delta(X)/k)/\varepsilon$
must be added to each centroid to obtain a
$\varepsilon$-differentially private output. 
 
For the above procedure to effectively reduce the noise added to the output
with respect to
 standard differential privacy
via noise addition with no prior microaggregation, 
the sensitivity $n/k\times \Delta(X)/k$ needs
to be smaller than the sensitivity of the data set without 
prior microaggregation, that is, $\Delta(X)$.
To that end, the $k$-anonymity parameter should be adjusted. 
Increasing $k$ has two effects: it reduces the contribution of each
record to the cluster centroid (thereby 
reducing the centroid sensitivity),
and it reduces the number of clusters (thereby
reducing the number of published centroids).
Specifically, for $n/k \times \Delta(X)/k$ to be less than
$\Delta(X)$, 
a value $k > \sqrt{n}$ is needed. This shows that the
utility benefits of this method  
depend on the size $n$ of the data set.

Here one must acknowledge that, while prior
microaggregation enables noise reduction as discussed 
above, microaggregating records into 
centroids also entails some utility loss. 
However, this loss is more
than compensated by the benefits brought by cluster
centroids being less sensitive than individual records.
This is so because microaggregation can exploit 
the underlying structure of data and reduce the 
sensitivity with relatively little utility loss. 
This hypothesis
is empirically supported by the extensive evaluations 
performed in~\cite{VLDB,TrustCom}. 

Even though this previous work effectively 
reduces the amount of Laplace noise to be added to reach
$\varepsilon$-differential privacy, the fact that
it requires using a microaggregation parameter $k$ that
depends on the number of records $n$ of the input data set may be problematic
for large data sets. 
In other words,
for large data sets, a value of $k$ so large may be required
that the utility loss incurred in the prior microaggregation step
cancels the utility gain due to subsequent noise reduction.

To circumvent this problem, in this paper we present an 
alternative procedure that
offers utility gains with respect to standard differential privacy 
regardless of the number of records of the input data set. 
Specifically, our method rests
on individual ranking univariate (rather than insensitive multivariate) microaggregation
to reduce sensitivity
in a way that only 
depends on the number of attributes to be protected. 
\footnote{Microaggregation is used to improve the utility of the released data, while privacy guarantees
	are provided by differential privacy alone. However, it is interesting to note that, while the multivariate
	microaggregation in~\cite{TrustCom,VLDB} yielded an intermediate $k$-anonymous data set, the individual ranking
	microaggregation in this paper yields an intermediate probabilistically $k$-anonymous~\cite{prob_k_anon} data set.}
This behavior is more desirable in at least the following cases:
i) data sets with a large number of records; ii) 
data sets with a small number of attributes; iii) 
data sets in which only the confidential attributes, which 
usually represent a small fraction of the total attributes, 
should be protected. 

\section{Differential privacy via individual ranking microaggregation}
\label{dp_indiv}

In this section we present a method to obtain 
differentially private data releases which, for 
specific data sets, may reduce noise even 
more than the above-mentioned 
approach based
on multivariate microaggregation.
First, we discuss in detail the limitations of that previous mechanism.
Then, we present a new proposal that, based on 
individual ranking,  
reduces the sensitivity of the microaggregated output independently 
of the number of records.
For simplicity, we first assume data sets with numerical attributes 
to which an amount
of Laplace noise is added to satisfy differential privacy. At the 
end of this 
section, we detail how the proposed method can be adapted to deal 
also with categorical attributes.

\subsection{Limitations of multivariate microaggregation for differentially private data releases}
\label{limit}

In~\cite{TrustCom,VLDB},
the utility gain was limited by the insensitive 
multivariate microaggregation.
Such microaggregation anonymizes the input data set at the record level.

The sensitivity of the set of $n/k$ centroids thus obtained 
is $n/k \times \Delta(X)/k$ because, in the worst case:
\begin{itemize}
\item Changing a single record in the input data set can
cause all $n/k$ clusters to change by one record
\item The record changed within each cluster can alter
the value of the cluster centroid by up to 
 $\Delta(X)/k$, where $\Delta(X)$ is the 
maximum distance between elements in the domain of the input data 
(we are assuming that centroids are computed as the arithmetic 
average of record values in the cluster).
\end{itemize}

The above worst-case scenario overestimates the actual 
sensitivity of the output and, thus, the noise to be added
to the centroids to achieve $\varepsilon$-differential privacy. 
Indeed, it is highly unlikely that 
modifying one input record by up to $\Delta(X)$
would change by $\Delta(X)$ one record in {\em each}
cluster. 
Let us consider an extreme scenario, 
in which all records in the input
data set take the \emph{maximum} possible value tuple in the domain of $X$. 
Recall that the insensitive microaggregation used 
sorts and groups records according to a total
order defined over the domain of $X$. 
Then, assume that the record located in the last position of the sorted list
changes to take the \emph{minimum} value tuple of the
domain of $X$, so that its distance to any of the other records in the data set is
$\Delta(X)$. According to the ordering criterion, such a change would 
cause the modified record to be 
``inserted'' in the first position of the sorted list.
Consequently,
all other records would be moved
to the next position, which would change \emph{all} 
clusters by one record. 
However, from the perspective of the centroid computation (\emph{i.e} the average of the record in the group), 
only the first cluster centroid,
where the modified record is located, would change and its variation
would be exactly $\Delta(X)/k$.

In other intermediate scenarios, the effect of modifying 
one record would be split among the centroids of the clusters
affected by the modification. 
Intuitively, 
the aggregation of the centroid variations would seem to be upper-bounded
by $\Delta(X)/k$, which
corresponds to the extreme case detailed above. However, this 
is only true if a total order for the domain of $X$
exists for which the triangular inequality is satisfied, that is, 
when $d(r_1,r_2)+d(r_2,r_3) \geq d(r_1,r_3)$ holds for any records $r_1$, $r_2$ and $r_3$ in $X$. 
Unfortunately, this is generally not the case
for multivariate data because a natural total order does not always exist.
Artificial total orders defined for multivariate data (for example, see the proposal in~\cite{VLDB})
do not fulfill the triangular inequality and, as discussed above, 
the sensitivity of
individual centroids should be multiplied by the number of released centroids ($n/k \times \Delta(X)/k$)
to satisfy differential privacy. 

The lack of a total order 
does not occur in univariate numerical data sets,
that is, those with just one attribute. 
With a single numerical attribute, 
a natural total order (the usual numerical order) can be easily
defined with respect to the minimum or maximum value of the domain of values
of the attribute so that the triangular inequality is fulfilled. 
In these conditions, it is shown in~\cite{Domi02} that clusters
in the optimal microaggregation partition contain consecutive
values. The next lemma shows that the sensitivity 
of the set of centroids is indeed $\Delta(X)/k$.

\begin{lemma}
\label{lem1}
Let $x_1, \cdots, x_n$ be a totally ordered set of values
that has been microaggregated into $\lfloor n/k \rfloor$ clusters
of $k$ consecutive values each, except perhaps one cluster
that contains up to $2k-1$ consecutive values. 
Let the centroids of these 
clusters be $\bar{x}_1, \cdots, \bar{x}_{\lfloor n/k \rfloor}$,
respectively. Now if, for any single $i$, $x_i$ is replaced
by $x'_i$ such that $|x'_i - x_i| \leq \Delta$ and new clusters
and centroids $\bar{x'}_1, \cdots, \bar{x'}_{\lfloor n/k \rfloor}$
are computed, it holds that
\[ \sum_{j=1}^{\lfloor n/k \rfloor} |\bar{x'}_j - \bar{x}_j| \leq \Delta/k.\]
\end{lemma}

{\bf Proof.} Assume without loss of generality that $x'_i > x_i$
(the proof for $x'_i < x_i$ is symmetric).
Assume, for the sake of simplicity,
that $n$ is a multiple of $k$ (we will later relax this 
assumption).
Hence, exactly $n/k$ clusters
are obtained, with cluster $j$ containing
consecutive values from $x_{(j-1)k+1}$ to $x_{jk}$. 
Let $j_i$ be the cluster to which $x_i$ belongs.
We can distinguish
two cases, namely $x'_i \leq x_{j_i k + 1}$ and $x'_i > x_{j_i k +1}$.

{\em Case 1}. 
When $x'_i \leq x_{j_i k +1}$, $x'_i$ stays in $j_i$. Thus, the
centroids of all clusters other than $j_i$ stay unchanged and the centroid
of cluster $j_i$ increases by $\Delta/k$, because
$x'_i = x_i + \Delta$.
So the lemma follows in this case.

{\em Case 2}. 
When $x'_i > x_{j_i k + 1}$, two or more clusters change as a result
of replacing $x_i$ by $x'_i$: cluster $j_i$ loses $x_i$ and another
cluster $j'_i$ (for $j'_i > j_i$) acquires $x'_i$.
To maintain its
cardinality $k$, after losing $x_i$, cluster $j_i$ acquires
$x_{j_i k+1}$. In turn, cluster $j_i + 1$ loses $x_{j_i k+1}$ and 
acquires $x_{(j_i+1) k+1}$, and so on, until cluster $j'_i$, 
which transfers its smallest value $x_{(j'_i-1) k +1}$ to 
cluster $j'_i -1$ and acquires $x'_i$. From cluster $j'_i+1$ upwards, nothing changes.
Hence the overall impact on centroids is  
\[ \sum_{j=1}^{n/k} |\bar{x'}_j - \bar{x}_j| 
= \sum_{j=j_i}^{j'_i} |\bar{x'}_j - \bar{x}_j| \]
\[ = \frac{x_{j_i k+1} - x_i}{k} + \frac{x_{(j_i+1)k+1} - x_{j_ik+1}}{k} + \cdots
+ \frac{x'_i - x_{(j'_i - 1) k +1}}{k} \]
\begin{equation}
\label{equ}
 = \frac{x'_i - x_i}{k} = \frac{\Delta}{k}.
\end{equation} 
Hence, the lemma follows also in this case.

Now consider the general situation in which $n$ is not a multiple 
of $k$. In this situation there are $\lfloor n/k \rfloor$ clusters
and one of them contains between $k+1$ and $2k-1$ values.
If we are in Case 1 above and 
this larger cluster is cluster $j_i$, the centroid of $j_i$ 
changes by less than $\Delta/k$, so the 
lemma also holds; of course if the larger cluster is one
of the other clusters, it is unaffected and the lemma also holds.
If we are in Case 2 above and the larger cluster is one 
the clusters that change, one of the fractions in 
the third term of Expression (\ref{equ})
above has denominator greater than $k$ and hence the overall
sum is less than $\Delta/k$, so the lemma also holds; if the 
larger cluster is one of the unaffected ones, the lemma also holds.
\hfill $\Box$

\subsection{Sensitivity reduction in multivariate data sets via individual ranking microaggregation}
\label{ranking}

From the previous section, 
it turns out that, for univariate data sets, 
the amount of noise needed to 
fulfill differential privacy after the microaggregation step is significantly 
lower than with the method in~\cite{VLDB} (\emph{i.e.} $\Delta(X)/k$ vs. $n/k \times \Delta(X)/k$).
Moreover, this noise is exactly $1/k$-th of 
the noise required by the standard differential privacy approach, 
in which the sensitivity is $\Delta(X)$ because
\emph{any} output record may change by $\Delta(X)$ 
following a modification of any record in the input
(as also stated in~\cite{Kellaris13} when sorting attributes by their value counts). 

To benefit from such a noise reduction in the case of 
multivariate data sets, 
we rely on the following composition property of differential privacy.

\begin{lemma}[Sequential composition \cite{McSherryPINQ}]
Let each sanitizing algorithm $Ag_i$ in a set of sanitizers provide 
$\varepsilon_i$-differential privacy. Then a sequence
of sanitizers $Ag_i$ applied to a data set $D$ provides
$(\sum_i \varepsilon_i)$-differential privacy.  
\end{lemma}

   In the context of differentially private data publishing, 
   we can think of a data release as the collected 
   answers to successive queries for each record in the data set.
   Let $I_r(X)$ be the query that returns the value of record $r$ (from a
   total of $n$ records) in the data set $X$. In turn, we can think of $I_r(X)$
   as the collected answers to successive queries for each of the attributes
   of record $r$. Let $I_{ra}(X)$ be the query function that returns the value
   of attribute $a$ (from a total of $m$ attributes). We have $I_r(X)=(I_{r1}(X),\ldots, I_{rm}(X))$.
   The differentially private data set that we seek can be generated by giving a
   differentially private answer to the set of queries $I_{ra}(X)$ for all $r=1,\ldots,n$ and
   all $a=1,\ldots,m$. Differential privacy being 
designed to protect the privacy of 
   individuals, such queries are very sensitive and require a large amount of noise.
   
   To reduce sensitivity and hence the amount of noise needed to attain
   differential privacy, we rely on individual ranking microaggregation 
   (which is more utility-preserving than multivariate microaggregation, 
   as explained in Section~\ref{micro} above). Instead of asking for 
$I_{ra}(X)$,
   the data set is generated by asking for individual ranking microaggregation centroids.
   Let $\rho_X(r,a)$ be the group of records of data set $X$ 
in the individual ranking 
   microaggregation of attribute $a$ that corresponds to $r$, and let $C_{\rho_X(r,a)}$
   be associated centroid. We replace $I_{ra}(X)$ by $C_{\rho_X(r,a)}$.
   
   Now, we have to minimize the amount of noise required to answer these queries in a 
   differentially private manner. We work with each attribute independently and
   then combine the queries corresponding to different attributes by applying sequential
   composition. If we get an $\varepsilon$-differentially private response to 
   $(C_{\rho_X(1,a)},\ldots,C_{\rho_X(n,a)})$ for each $a=1,\ldots,m$, then joining
   them we have $m\varepsilon$-differential privacy.
   
   For attribute $a$, we have to answer the query  $(C_{\rho_X(1,a)},\ldots,C_{\rho_X(n,a)})$
   in an $\varepsilon$-differentially private manner. 
If we compute the $L_1$-sensitivity of this
   query, $s_a$, we can attain $\varepsilon$-differential privacy by adding a Laplace distributed
   noise with scale parameter $s_a/\varepsilon$ to each component
$C_{\rho_X(i,a)}$. 
   We have already seen that for individual ranking microaggregation the $L_1$-sensitivity
   of the list of centroids is $\Delta_a/k$. However, in our query each centroid appears
   $k$ (or more times); hence, the sensitivity is multiplied 
by $k$ and becomes $\Delta_a$ (or greater),
   which is not satisfactory. Our goal is to show that we can attain $\varepsilon$-differential privacy
   by adding a Laplace noise with scale $\Delta_a/(k\varepsilon)$ rather than $\Delta_a/\varepsilon$ (as 
   an $L_1$-sensitivity of $\Delta_a$ would require). To that end, 
instead of taking an independent draw 
   of the noise distribution for each of the components, we use the same draw for all the components that
   refer to the same centroid. That is, we use the same random variable 
$L_{\rho_X(r,a)}$ to determine the
   amount of noise that is added to all the components sharing
the same value $C_{\rho_X(r,a)}$;
similarly, in data set $X'$ 
we use $L_{\rho_{X'}(r,a)}$ as noise for all components
 sharing the same value $C_{\rho_{X'}(r,a)}$.
   For $\varepsilon$-differential privacy, it must be
   \[
   \frac{\Pr( (C_{\rho_X(1,a)}+L_{\rho_X(1,a)},\ldots,C_{\rho_X(n,a)}+L_{\rho_X(n,a)})  = (x_1,\ldots,x_n)) }
        {\Pr( (C_{\rho_{X'}(1,a)}+L_{\rho_{X'}(1,a)},\ldots,C_{\rho_{X'}(n,a)}+L_{\rho_{X'}(n,a)}) = (x_1,\ldots,x_n))}
   \le 
   \exp(\varepsilon).
   \]
   
If any of $x_1,\ldots, x_n$ is not a centroid value plus the noise
corresponding to that centroid value (note that equal centroid
values are added equal noise values, as said above), 
the probabilities in both the numerator and the denominator
of the above expression are zero, and differential privacy is satisfied.
Otherwise, we have that $x_1,\ldots, x_n$ are only repetitions of 
$n/k$ different values, that is, the values of the $n/k$ centroids
plus the noise corresponding to each centroid value.
Thus,  we can simplify the expression by removing 
all but one of each of those repetitions. Let
   $C_{i,a}(X)$ and $C_{i,a}(X')$ for $i=1,\ldots,n/k$ be the centroid 
values for attribute $a$ associated to $X$ and $X'$, respectively, and
   $L_{i,a}$ and $L'_{i,a}$ be 
Laplace noises with scale $\Delta_a/(k\varepsilon)$ associated to 
those centroid values, respectively. 
After rewriting the above inequality
   in these terms and taking into account that the sensitivity of the list of centroids is $\Delta_a/k$, it is
   evident that $\varepsilon$-differential privacy is satisfied.
   \[
   \frac{\Pr( (C_{1,a}(X)+L_{1,a},\ldots,C_{n/k,a}(X)+L_{n/k,a})  = (x'_1,\ldots,x'_{n/k})) }
   {\Pr( (C_{1,a}(X')+L'_{1,a},\ldots,C_{n/k,a}(X')+L'_{n/k,a})  = (x'_1,\ldots,x'_{n/k}))}
   \le 
   \exp(\varepsilon).
   \]

Hence, we propose the following algorithm to obtain 
a differentially private version $X^D$ of a numerical original data set $X$
with attributes $A_1, \cdots, A_m$.

\begin{algorisme}
\label{alg1}~\\
\begin{enumerate}
\item\label{pas1} Use individual-ranking microaggregation independently
on each attribute $A_i$, for $i=1$ to $m$.
Within each cluster,
{\em all} attribute values are replaced by 
the cluster centroid value, so each microaggregated cluster
consists of $k$ repeated centroid values. 
Let the resulting microaggregated
data set be $X^M$.
\item\label{pas2} Add Laplace noise independently to each attribute 
$A^M_i$ of $X^M$,
where the scale parameter for attribute $A^M_i$ is 
\[\Delta(A^M_i)/\varepsilon = \Delta(A_i)/(k\times\varepsilon).\]
The {\em same} noise perturbation is used on all 
repeated centroid values within each cluster.  
\end{enumerate}
\end{algorisme}

Now we can state:

\begin{lemma}
\label{lem2}
The data set  output by Algorithm~\ref{alg1} is
$m \varepsilon$-differentially private.
\end{lemma}

 {\bf Proof}. The lemma follows from the discussion preceding Algorithm~\ref{alg1}. 
  \hfill $\Box$

{\bf Note.} In Step~\ref{pas2} of Algorithm~\ref{alg1}, it is critically 
important to add exactly the same noise perturbation to 
all repeated values within a microaggregated cluster.
If we used different random perturbations for each repeated 
value, the resulting noise-added cluster
would be equivalent
to the answers to $k$ independent queries. This would
multiply by $k$ the sensitivity of the centroid,  
which would cancel the 
sensitivity reduction brought by microaggregation in Step~\ref{pas1}. 

\subsection{Choosing the microggregation parameter $k$}
\label{param}

In order to obtain an $\varepsilon$-differentially private
data set, by parallel composition
it suffices to make each 
record $\varepsilon$-differentially
private. In turn, to make a record $\varepsilon$-differentially
private, we have
two possibilities:
\begin{enumerate}
\item {\em Plain Laplace noise addition without microaggregation}.
Given that each record has $m$ attributes, by sequential composition
we need $(\varepsilon/m)$-differentially private attribute
values to obtain an $\varepsilon$-differentially private record. 
Hence, Laplace noise addition with scale 
parameter $\Delta(A_i)/(\varepsilon/m)=m\Delta(A_i)/\varepsilon$
needs to be added to each attribute $A_i$. 
\item {\em Our approach}. When performing 
individual-ranking microaggregation and replacing original values by cluster 
centroids, we preserve the structure of records.
By sequential composition, to make a record of $X^M$  $\varepsilon$-differentially private,
we need to make attributes in $X^M$ 
$(\varepsilon/m)$-differentially private.  
Hence, Laplace noise addition with scale
parameter $\Delta(A^M_i)/(\varepsilon/m)=m\Delta(A^M_i)/\varepsilon$ 
needs to be added to each attribute $A^M_i$. However, dealing
with $A^M_i$ rather than $A_i$ is better, because $A^M_i$ is less
sensitive. Indeed, $\Delta(A^M_i)=\Delta(A_i)/k$, so the scale
parameter is $m\Delta(A^M_i)/(k\varepsilon)$.
\end{enumerate}

According to the above discussion, our approach adds less noise
than plain Laplace noise addition for any $k > 1$.
Admittedly, its prior individual
ranking microaggregation causes some additional information loss. 
However, this information 
loss grows very slowly with the cluster size $k$ and also
with the number of attributes $m$ (see~\cite{Domi01}),  
whereas the Laplace noise being added decreases very quickly with $k$.
The experiments in Section~\ref{evaluation} below show
that the information loss caused by individual ranking
is negligible in front of Laplace noise addition.

\subsection{Dealing with categorical attributes}
\label{categ}

So far, we have assumed that attributes in the input data set are numerical,
so that: i) they are totally ordered;
ii) centroids can be computed as standard numerical averages; 
and iii) Laplace noise with an appropriate scale parameter 
can be added to satisfy differential privacy.
However, many data sets contain attributes with categorical values, 
such as Ethnicity, Country of birth or Job. 
Unlike numerical attributes,
categorical attributes take values from a finite set of
categories for which the arithmetical operations needed
to microaggregate and add noise to the outputs 
do not make sense.
Following the discussions in~\cite{VLDB}, 
some alternative mechanisms can be used to adapt the above-described method to
categorical attributes:
\begin{itemize}
\item Unlike numbers, the domain of values of a categorical attribute should be defined by extension.
Ways to do this are a flat list or a hierarchy/taxonomy.
The latter is more desirable, 
since the taxonomy implicitly captures the semantics inherent to 
categorical values ({\em e.g.}, disease categories, 
job categories, sports categories, etc.). In this manner, 
further operations can exploit this 
taxonomic knowledge and provide a semantically coherent management of 
attribute values, which is usually the most important 
dimension of utility for categorical data~\cite{Martinez12a}.
\item A suitable function is needed to compare categorical values 
that exploits the semantics provided by the corresponding taxonomies
(if any). Semantic distance measures quantify the amount 
of differences observed between two categorical values according to the knowledge modeled in 
a taxonomy. 
In~\cite{VLDB} several measures available 
in the literature
were discussed from the perspective of differential privacy,
and the measure proposed
in~\cite{Sanc12} was evaluated as the most suitable one. It 
computes the distance $\delta: A_i\times A_i\rightarrow\mathbb{R}$
between two categorical values $a^i_1$ and $a^i_2$ of attribute $A_i$, 
whose domain is modeled in the taxonomy $\tau(A_i)$,
as a logarithmic function of their number of non-common taxonomic ancestors divided (for normalization)
by their total number of ancestors:

\begin{equation}
\delta(a^i_1, a^i_2) = \log_2 \Bigg(1+ \frac{|\phi(a^i_1) \cup \phi(a^i_2)|-|\phi(a^i_1) \cap \phi(a^i_2)|}{|\phi(a^i_1) \cup \phi(a^i_2)|} \Bigg), \label{eq:d}
\end{equation}

where $\phi(a^i_j)$ is the set of taxonomic ancestors of $a^i_j$ in $\tau(A_i)$, including itself.

The advantages of this measure are: i) it captures subtle differences between values modeled in the taxonomy
because all taxonomic ancestors are considered;
ii) thanks to its non-linearity, its sensitivity to 
outlying values is low, 
which is desirable to reduce the sensitivity of data; 
iii) it fulfills the following measure properties: \textit{non-negativity}, 
\textit{reflexivity}, \textit{symmetry} and 
\textit{triangle inequality}, thereby defining 
a coherent total order within the attribute categories.
\item A total order that yields insensitive and within-cluster
homogeneous microaggregation can be defined through the
notion of \textit{marginality}~\cite{infsci2013}. Marginality
measures how far 
each categorical value of the attribute in the data set
lies from the ``center'' of that attribute's taxonomy, 
according to a semantic distance (like the above-described one).
A total order between categorical values can be defined based
on their marginality: categorical values present in the data set
are sorted according to their distance to the most marginal value
(\emph{i.e.} the categorical value farthest from the center of the domain).
The {\em marginality $m(\cdot, \cdot)$} of each value $a^i_j$ in $A_i$ 
with respect to its domain of values $Dom(A_i)$ is computed as
\begin{equation}
m(Dom(A_i), a^i_j) = \sum_{a^i_l \in Dom(A_i) -\{a^i_j\}} \delta(a^i_l,a^i_j)   \label{eq:m}
\end{equation}
where $\delta(\cdot,\cdot)$ 
is the semantic distance between two values.
The grea\-ter $m(Dom(A_i), a^i_j)$, the more marginal 
({\em i.e.}, the less central) is $a^i_j$ with regard to $Dom(A_i)$.

\item Microaggregation replaces original values in a cluster 
by the cluster centroid,
which is the arithmetical mean in case of numerical data.
Centroids for categorical data can be obtained by relying again
on marginality: the mean of 
a sample of categorical values can be
approximated by the least marginal value in the taxonomy, that is, the
value that minimizes the aggregated distances to all other elements in the data set~\cite{Martinez12}.
Formally, given a sample $S(A_i)$ of a nominal attribute $A_i$ in 
a certain cluster, 
the marginality-based centroid for that 
cluster is defined as
\begin{equation}
Centroid(S(A_i)) = \arg\min_{a^i_j \in \tau(S(A_i))}m(S(A_i),a^i_j), \label{eq:cen}
\end{equation}
where $\tau(S(A_i))$ is the minimum taxonomy extracted 
from $\tau(A_i)$ that
includes all values in $S(A_i)$.
\item Finally, to satisfy differential privacy, an amount 
of uncertainty proportional
to each attribute's sensitivity should be added prior to releasing the data. 
Since adding Laplace noise to categorical centroids makes no sense, an alternative way to
obtain differentially private outputs consists in selecting attribute centroids in a 
probabilistic manner. This can be done by means 
of the Exponential Mechanism proposed in~\cite{McSherry}. This mechanism 
chooses the centroid closest to the optimum 
(\emph{i.e.} least marginal value, in our case) 
according to the input data, the $\varepsilon$-differential privacy parameter
and a quality criterion, which in this case is the marginality of each categorical value. 
Formally, given a function with discrete outputs $t$, 
the mechanism chooses an output that is
close to the optimum according to the input data $X$ 
and quality criterion $q(X,t)$, while preserving $\varepsilon$-differential privacy. 
Each output is associated with a selection probability
$\Pr(t)$, which grows exponentially with the quality criterion, as follows:
\[\Pr(t) \propto \exp(\frac{{\varepsilon}q(X,t)}{2\Delta(q)}) \]
Algorithm~\ref{alg:catdif} describes the application of this mechanism
to select $\varepsilon$-differentially private centroids.
\end{itemize}

\begin{algorisme}
\label{alg:catdif}~\\

{\bf let} $C$ be a cluster with at least $k$ values of the attribute $A_i$
\begin{enumerate}
\item\label{p1}Take as the quality criterion $q(\cdot,\cdot)$ for
each centroid candidate $a^i_j$ in $\tau(A_i)$ the additive 
inverse of its marginality towards the attribute values $S(A_i)$ contained 
in $C$, that is, $-m(S(A_i),a^i_j)$; 

\item\label{p2}Sample the centroid from a distribution 
that assigns
\begin{equation}
\label{eq:exp}
\Pr(a^i_j) \propto \exp(\frac{{\varepsilon} \times (-m(S(A_i),a^i_j))}{{2}\Delta(m(A_i))}).
\end{equation}

\end{enumerate}
\end{algorisme}

\section{Empirical evaluation}
\label{evaluation}

This section details the empirical evaluation
of the proposed method (in terms of noise reduction and utility preservation)
in comparison with the method in~\cite{VLDB}. 

\subsection{Evaluation data}

As evaluation data we used two data sets:
\begin{itemize}
\item ``Census''~\cite{Brand},
which is a standard data set meant for testing privacy protection methods.
It was used in the European project CASC and 
in~\cite{Dand02b,Yanc02,Lasz05,Domi05,Domingo10}
and it contains 1,080 records with 13 numerical attributes. 
Since all attributes represent 
non-negative numerical magnitudes (\emph{i.e.} money amounts), we 
defined the domains of the attributes as $[0 \ldots (1.5 \times max\_attr.\_value\_in\_data set)]$. 
The domain upper bound is a 
reasonable estimate if the attribute values in the data set are
representative of the attribute values in the population, 
which in particular
means that the population outliers are represented in the data set.
The difference between the bounds of the domain of each attribute $A_i$ 
determines the sensitivity of that attribute ($\Delta(A_i)$) 
and, as detailed above,
determines the amount of Laplace noise to be added to microaggregated outputs.
Since the Laplace distribution takes
values in the range $(-\infty, +\infty)$, 
for consistency we bound noise-added outputs to the domain ranges define above.
\item ``Adult''~\cite{Adult}, which is a well-known data set from the UCI repository.
It contains both numeric and categorical attributes and it 
was also used in~\cite{VLDB}.
To enable a fair comparison, we used the same attributes as in~\cite{VLDB}.
As categorical attributes, we used OCCUPATION 
(that covers 14 distinct categories)
and NATIVE-COUNTRY (with 41 categories). The taxono\-mies 
modeling attribute domains 
were extracted from WordNet 2.1~\cite{Fellbaum98}, a general-purpose repository that taxonomically
models more than 100,000 concepts. Mappings between 
attribute labels and WordNet concepts
are those stated in~\cite{Martinez12a}. 
Domain boundaries for each attribute and
sensitivities for centroid quality criteria were set as described
in Section~\ref{categ}.
As numeric attributes, we used AGE and (working) HOURS-PER-WEEK.
Domain boundaries and sensitivities for these two numerical attributes were computed
as explained for ``Census''.
The experiments were carried out with the training 
corpus from the Adult data set, which 
consists of 30,162 records after removing records with missing values of the considered attributes.
\end{itemize}

\subsection{Evaluation measures and experiments}
\label{sec:eval1}

Since our proposal aims at providing differentially private outputs
while making as few assumptions
on their uses as $k$-anonymity-like models do, we 
used generic utility metrics, as usual in the 
literature on $k$-anonymity and statistical disclosure control
({\em e.g.}~\cite{Domi05}).
In that literature, the utility 
of the anonymized output is evaluated in terms of \textit{information loss}.
Information loss measures the differences
between the original and the anonymized data set.  

A well-known generic information loss measure 
that is suitable to capture the distortion of the output is 
the Relative Error (RE), which measures the error of 
an answer as a function of the answer's magnitude;
in this manner, answers with large magnitudes can tolerate larger errors~\cite{Xiao11}. 
In our setting, for each numerical attribute value $a_{j}^{i}$, RE is measured 
as the absolute difference between
each original attribute value $a_{j}^{i}$ and its masked
 version $(a_{j}^{i})'$ divided by the original value.
\[ Relative\_Error(a_{j}^{i}) = \frac{|a_{j}^{i}-(a_{j}^{i})'|}{\max(sanity\_bound(A_i), |a_{j}^{i}|)} \]

A sanity bound is included in the denominator to 
mitigate the effect of excessively small values.
Since in our approach we aim at publishing the attribute
values of the records in the data set, we defined this
sanity bound as a function of the domain of values $Dom(A_i)$
each attribute $A_i$; specifically, $sanity_bound(A_i)=Dom(A_i)/100$.
In this manner, the mitigation effect of the sanity
bound is adapted to the order of magnitude of each attribute.

For categorical attributes we directly measured the error as the 
semantic distance (Expression (~\ref{eq:d})) between original 
and masked records,
which is already normalized in the 0..1 range. 
The Relative Error of the whole data set $X$ is measured as the
average Relative Error of all attributes of all records.
Notice that with a high error, that is, a high information loss, 
a lot of data uses are severely damaged, like
for example subdomain analyses (analyses restricted to parts of the data set). 

Another generic information loss measure focuses 
on how different are variances of attributes in the
original and anonymized data sets~\cite{Baeyens99}. 
Preserving the sample variance is relevant for statistical
analysis and closely depends on how constrained 
the microggregation algorithm is.
The variation of the attribute variances (for numerical attributes) has been measured as follows:
\[ \Delta(\sigma^2(A_{i})) = \frac{\mid\sigma^2(A'_{i})-\sigma^2(A_{i})\mid}{\mid\sigma^2(A_{i})\mid}, \]
where $\sigma^2(A_{i})$ and $\sigma^2(A'_{i})$ denote the variances of attribute $A_{i}$ 
in the original data set and its masked version, respectively.

Moreover, since many works on differential privacy
focus on preserving the utility of counting queries
~\cite{Xiao2010,Xiao2010b,Xu2012,Blum,Dwor09,Hardt2010,Chen11},
we measured how the methods preserve the data distribution by 
building histograms of each attribute and comparing the distribution
between the original and masked values according to the well-known
Jensen-Shannon divergence (JSD)~\cite{Lin91}, which is symmetric and bounded
in the 0..1 range.  
At a data set level, we averaged the divergence of all the attributes.
Histograms for continuous attributes (i.e., those of the ``Census''
data set) have been created by grouping records in bins accounting
1/100th of the attribute domain. In the ``Adult'' data set, in which
all the attributes (either numeric or categorical) are discrete and have a limited set of values,
each bin corresponds to one individual attribute value.

The $\varepsilon$ parameter for differential privacy was set to 
$\varepsilon= \{0.1, 1, 10\}$, which covers the usual range of
differential privacy levels observed in the 
literature~\cite{Dwork11,Char10,Char12,Mach08,VLDB}.
As discussed in Section~\ref{ranking}, the scale parameter
of the Laplace noise needed by our method 
to achieve $\varepsilon$-differential 
privacy is $\Delta(A_i)/(k\times(\varepsilon/m))$;
that is, it depends on the level $k$ of prior microaggregation 
and on the number $m$ of attributes to be protected.  
We evaluated the influence of these parameters in the ``Census''
data set by taking $k$ between 2 and 100 and $m \in \{1, 8, 13\}$.
For the ``Adult'' data set, as done in ~\cite{VLDB},
we set $k$ between 2 and 200.

\subsection{Comparison with baseline methods}
\label{sec:eval2}

In order to benchmark the results of our proposal, we
considered the following baseline methods:
\begin{itemize}
\item Plain Laplace noise addition 
for $\varepsilon$-differential privacy
as described in Section~\ref{param}
above. 
Even though this mechanism is the naivest way to produce 
differentially private data sets, it is useful as an 
upper bound of information loss. By comparing against it, 
we can quantify
the gain brought by the prior microaggregation steps.
\item Plain individual ranking, with no subsequent
Laplace noise addition. Although this method does not 
lead to $\varepsilon$-differential privacy by itself, we want
to show the contribution 
of individual ranking to the information loss caused by our
method.
\end{itemize}

\begin{figure*}
\begin{center}
\includegraphics[width=12.5cm]{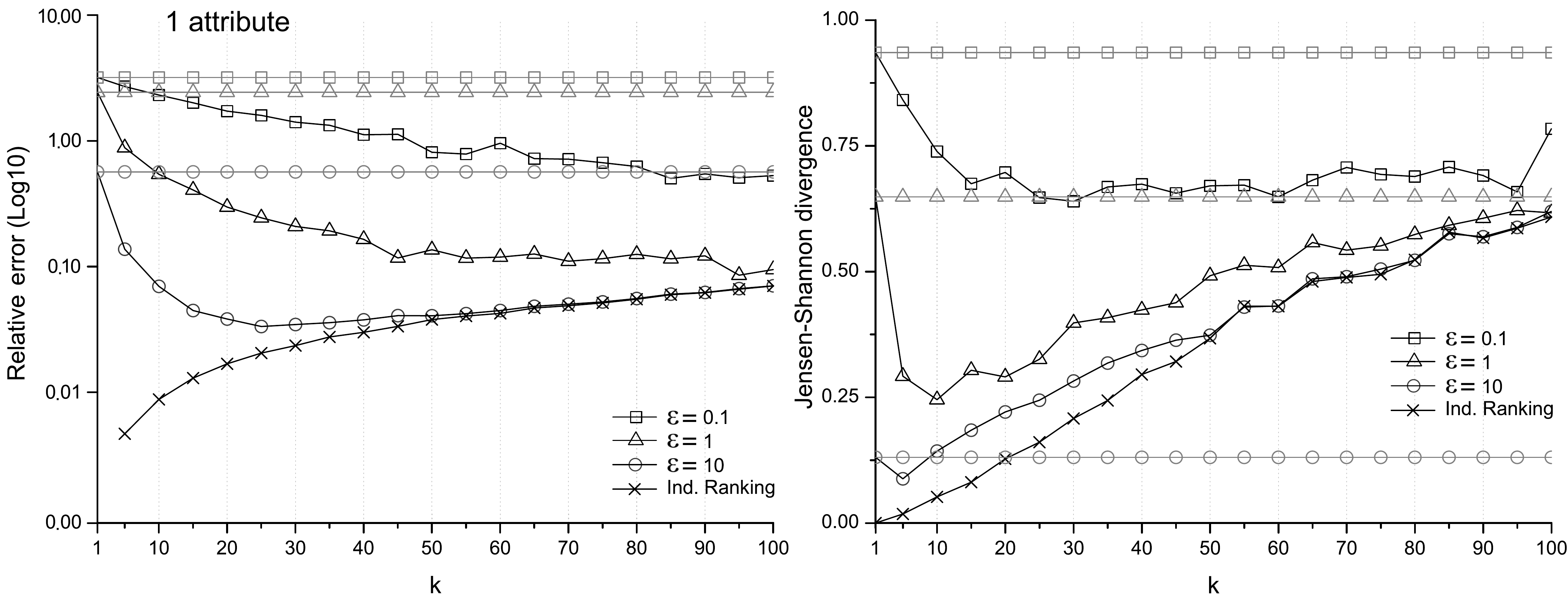}\\
	\vspace{0.2cm}
\includegraphics[width=12.5cm]{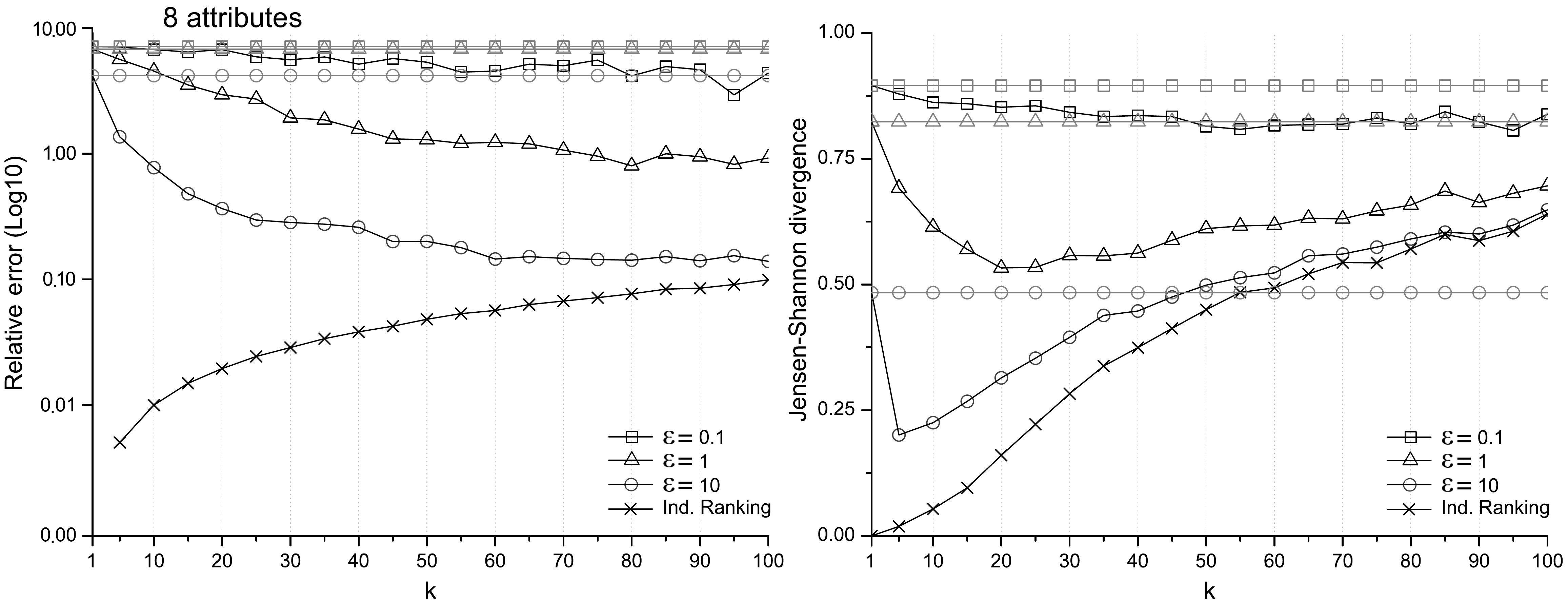}\\
	\vspace{0.2cm}
\includegraphics[width=12.5cm]{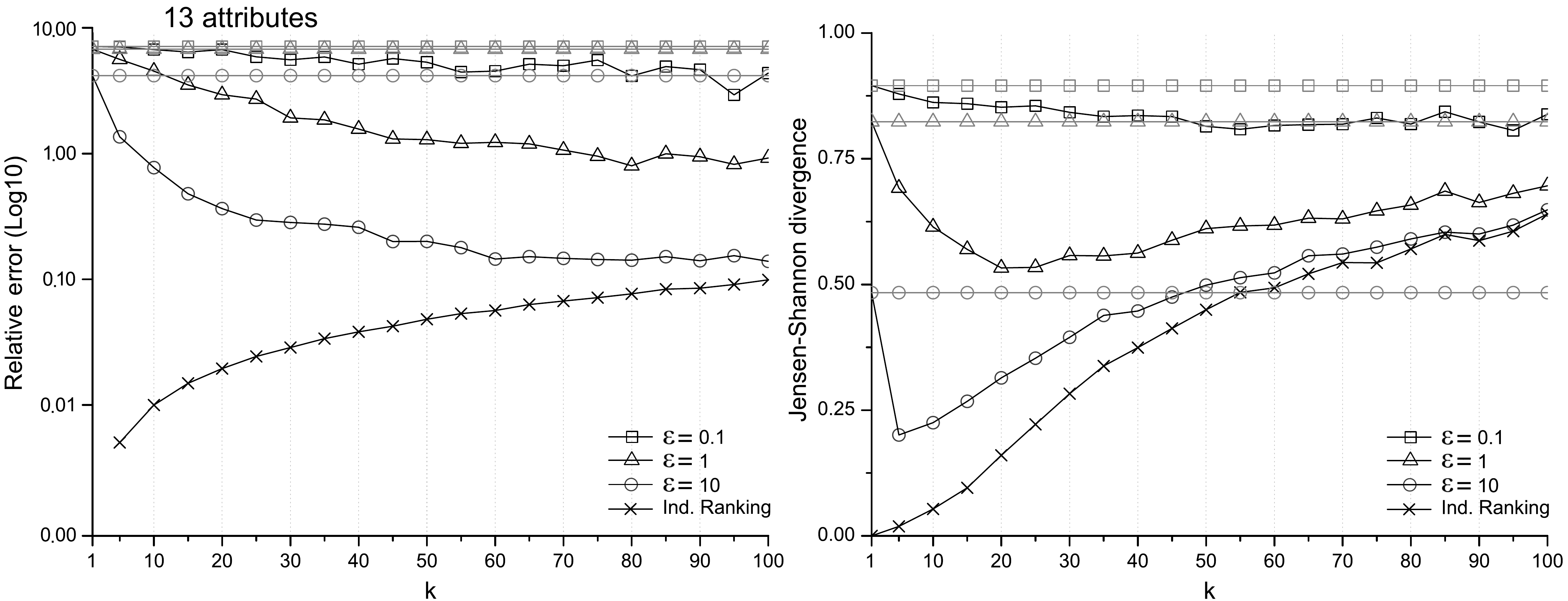}\\
	\vspace{0.2cm}
\caption{``Census'' data set: RE (on the left) and JSD (on the right) for the proposed method
for several numbers of attributes ($m$) and
$\varepsilon$ values (black non-horizontal lines, as RE and JSD
depend on the microaggregation
parameter $k$) compared to plain Laplace noise addition 
(gray horizontal lines,
because RE and JSD do not depend on $k$) and plain individual ranking
microaggregation.}
\label{figIRDP_Census}
\end{center}
\end{figure*}

\begin{figure*}
\begin{center}
\includegraphics[width=12.5cm]{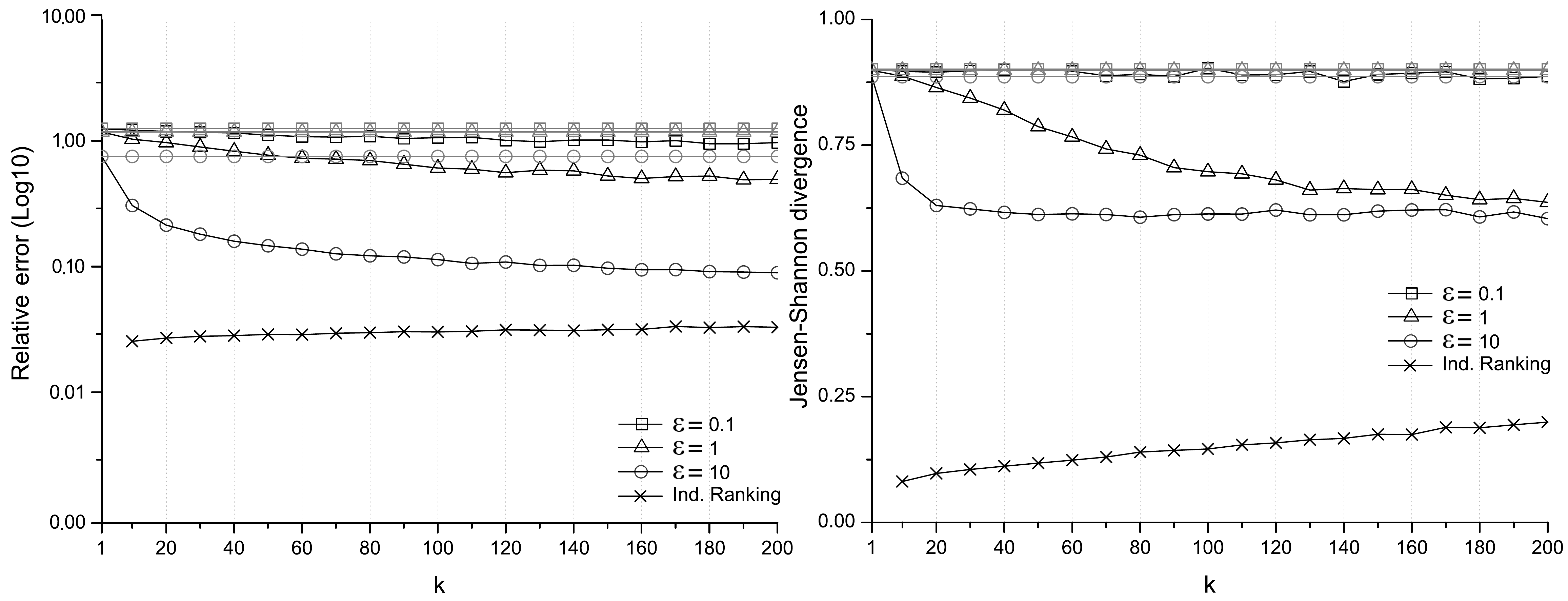}
\caption{``Adult'' data set: RE (on the left) and JSD (on the right) for the proposed method
for several $\varepsilon$ values (black non-horizontal lines, as RE and JSD
depend on the microaggregation
parameter $k$) compared to plain Laplace noise addition 
(gray horizontal lines,
because RE and JSD do not depend on $k$) and plain individual ranking
microaggregation.}
\label{figIRDP_Adult}
\end{center}
\end{figure*}

We computed RE and JSD for the baseline methods above and 
our approach for the two evaluation data sets.
Figure~\ref{figIRDP_Census} shows the comparison between
plain Laplace noise addition,
plain individual ranking and our approach for ``Census'';
Figure~\ref{figIRDP_Adult} shows 
the same comparison for ``Adult''.
Due to the broad ranges of the RE values,
a $\log_{10}$ scale is used for the Y-axes.

The plain Laplace noise addition baselines 
are displayed as gray horizontal lines, because they do not
depend on the value of $k$.
Each test involving Laplace noise shows 
the average results of 10 runs,
for the sake of stability. In any case, the
spikes shown in the graphs are the result of the randomization
inherent to Laplace noise addition.

Both Figure~\ref{figIRDP_Census} and Figure~\ref{figIRDP_Adult} 
show that, already for $k>1$, our approach reduces
the noise required to attain $\varepsilon$-differential privacy
compared to plain Laplace noise addition. This confirms
what was said in Section~\ref{param}.

The relative improvement of RE depends on the value of $\varepsilon$.
For $\varepsilon=0.1$ the amount of
noise involved is so high that even with the noise
reduction achieved by our method, the output data are hardly useful.
For $\varepsilon=10$ there is a 
substantial decline of RE for low $k$, whereas for larger $k$
RE stays nearly constant and is almost as low as   
the RE achieved by individual ranking alone.
This is especially noticeable when only one attribute is protected
(for the ``Census'' data set) for which the information loss 
of the proposed method even grows for $k > 25$; with a single attribute, 
we do not need to apply
sequential composition (that would require more noise to be
added) in order to attain a certain level of $\varepsilon$-differential 
privacy; at the same time, while for large $\varepsilon$ 
the RE of the method can be 
as low as the RE of individual ranking, it cannot be lower, which
explains why the former RE grows for large $k$ in the top-left graph of 
Figure~\ref{figIRDP_Census} (in that graph the RE of the method
becomes equal to the RE of individual ranking and the latter RE 
grows with $k$). 
Thus, for large $\varepsilon$,
the distortion added by individual ranking in larger clusters limits
the effect of the noise reduction achieved 
at the $\varepsilon$-differential privacy stage
due to the decreased sensitivity with larger $k$.
For the ``Adult'' data set, the difference in RE between the 
$\varepsilon$-differentially private outcome and the plain 
individual ranking is more noticeable because of the need to
discretize noise-added values.
 However, the larger cardinality
of the data set also allows using larger $k$ values
(to reduce the sensitivity) than in ``Census''
with comparable utility damage.

The comparison of distributions between the original and the masked data 
also outputs a similar pattern: JSD improves for $k>1$.
For the ``Adult'' data set,
Figure~\ref{figIRDP_Adult}
shows that the behaviors of JSD and RE are similar. 
For the ``Census'' data set, however, Figure~\ref{figIRDP_Census}
reports a sharp decline of JSD for low
$k$, whereas for large $k$ the divergence of distributions tends to 
increase; 
this is especially noticeable for $\varepsilon=10$,
whose results for $k>10$ (for a single attribute)
and for $k>45$ (for multiple attributes) are even worse than
with with plain Laplace noise. In these cases, the distortion introduced
by the individual ranking microaggregation dominates 
the gain brought by the reduced noise addition. 
This is clearer when the number
of attributes is small, because so is the noise to be 
added to fulfill $\varepsilon$-differential privacy. 
It is important to note that, for the ``Census'' data set,
the continuous values of the attributes have been discretized
in bins covering 1/100th of the attribute domain. Thus,
a microaggregation with a low $k$ would tend to cluster
values that fall within the same bin, which explains
the low distortion incurred in the (discretized) distributions.
For larger $k$, however, microaggregation tends to 
group records of different bins, which significantly 
alters the data distribution; this is precisely what happens for the
``Adult'' data set for any value of $k$, because bins cover individual
attribute values.

\subsection{Comparison with prior multivariate microaggregation}
\label{sec:eval3}

In a second battery of experiments, we compared the proposed method with
the previous work~\cite{VLDB}, in which records were
microaggregated using an insensitive version of the MDAV 
multivariate algorithm~\cite{Domi05}.

\begin{figure*}
\begin{center}
\includegraphics[width=12.5cm]{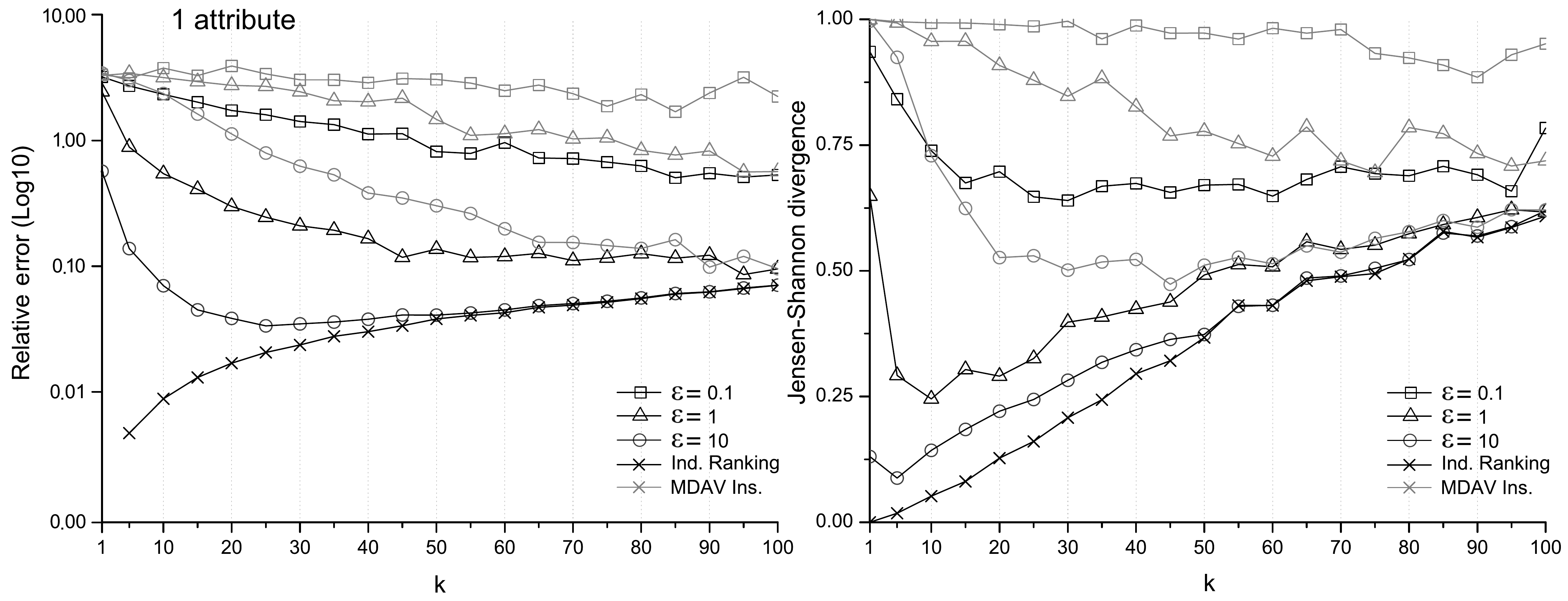}\\
	\vspace{0.2cm}
\includegraphics[width=12.5cm]{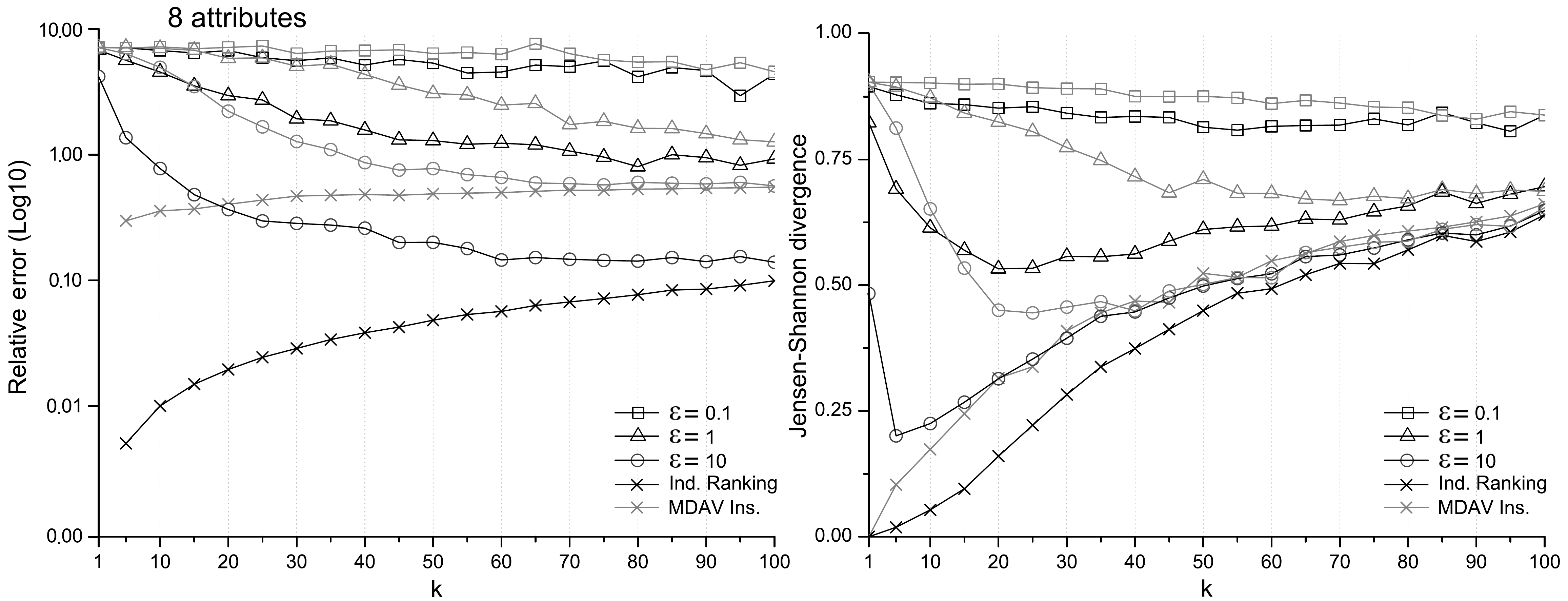}\\
	\vspace{0.2cm}
\includegraphics[width=12.5cm]{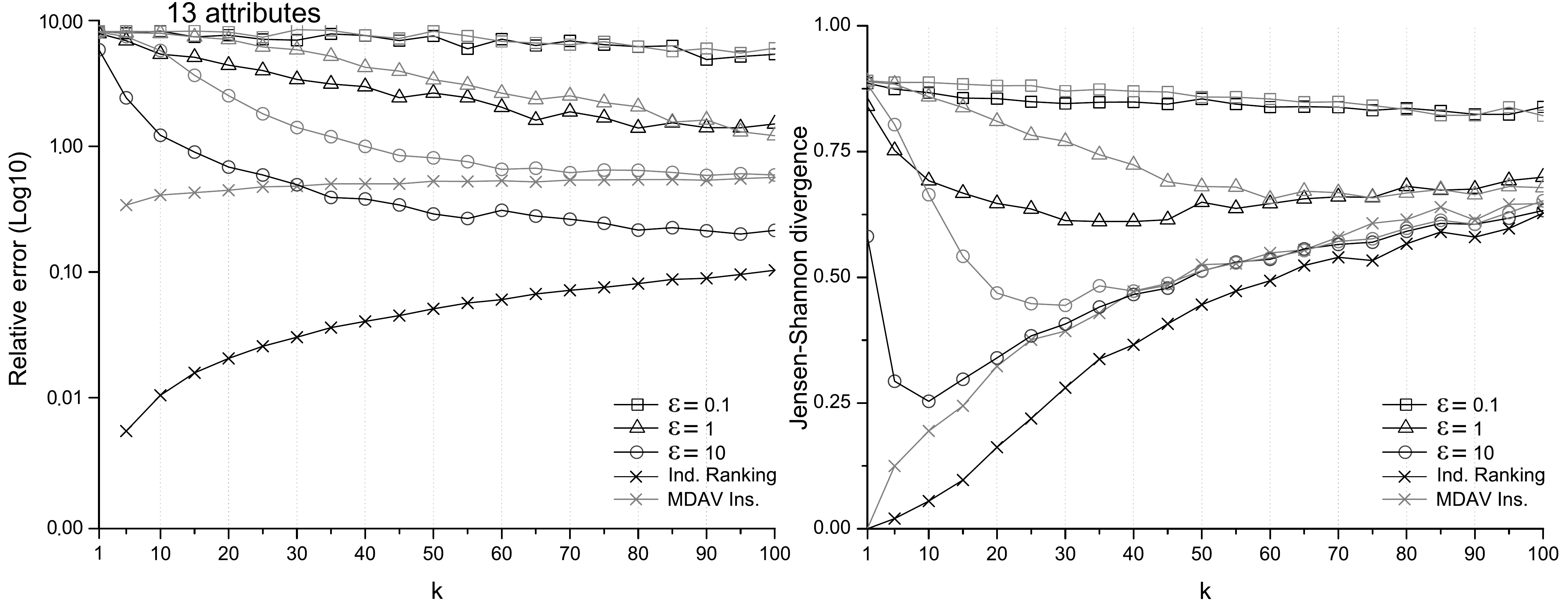}\\
	\vspace{0.2cm}
\caption{``Census'' data set: RE (on the left) and JSD (on the right) in the proposed method for several numbers 
of attributes ($m$) and 
$\varepsilon$ values (black dashed lines) compared to the previous work~\cite{VLDB} 
(gray dashed lines),
the insensitive version of the MDAV microaggregation algorithm
(gray solid lines) and the 
plain individual ranking method (black solid lines).}
\label{figIRkDP-Census}
\end{center}
\end{figure*}

\begin{figure*}
\begin{center}
\includegraphics[width=12.5cm]{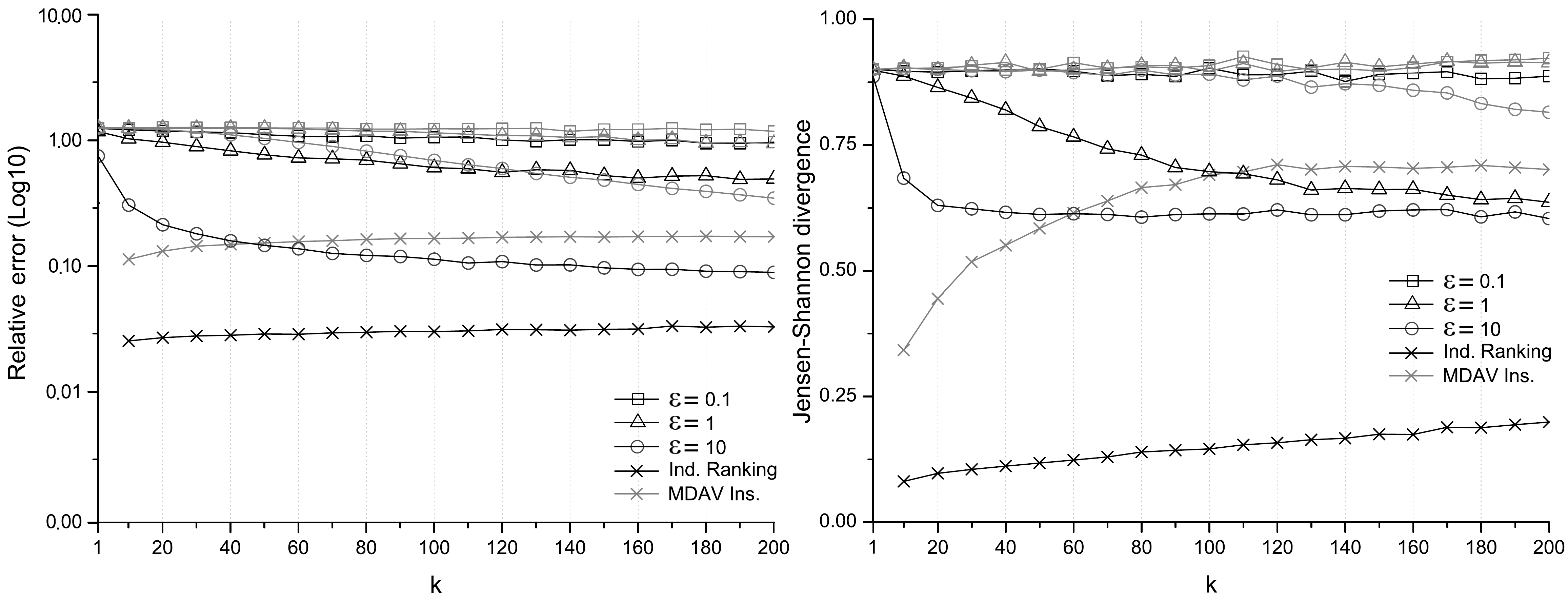}\\
\caption{``Adult'' data set: RE (on the left) and JSD (on the right) in the proposed method for several
$\varepsilon$ values (black dashed lines) compared to the previous work~\cite{VLDB} 
(gray dashed lines),
the insensitive version of the MDAV microaggregation algorithm
(gray solid lines) and the 
plain individual ranking method (black solid lines).}
\label{figIRkDP-Adult}
\end{center}
\end{figure*}

Similarly to previous figures, Figures~\ref{figIRkDP-Census} 
and~\ref{figIRkDP-Adult} depict the RE and JSD values 
for the different parameterizations of $k$, $\varepsilon$ and $m$,
for the ``Census'' and ``Adult'' data sets, respectively.
The results of the proposed method are represented with black lines whereas
the results of the previous work~\cite{VLDB} are displayed 
by gray lines with the same pattern for each $\varepsilon$ value.
As baselines, we also added the RE and JSD values 
incurred by the individual ranking
and the insensitive MDAV microaggregation algorithms alone. 

First, we notice that insensitive MDAV multivariate microaggregation 
alone incurs a
noticeably higher RE than plain individual ranking
microaggregation for $m > 1$. 
JSD is also higher for ``Census''
and significantly higher for ``Adult'',
because in the latter case the already discrete
values are not grouped into bins and, thus,
any change is reflected as a divergence in distributions.
In fact, for ``Adult'' our differentially private
method is even able to improve the figures of insensitive MDAV
for $k>100$.
For $m = 1$, both types of microaggregation 
are equivalent, so their REs and JSDs overlap.
On the other hand, as discussed in Section~\ref{param},
for multivariate data sets ($m > 1$), individual ranking 
microaggregation yields more homogeneous clusters
than MDAV multivariate microaggregation, because the 
former method builds clusters 
on an attribute-by-attribute basis, whereas the latter
builds clusters by taking all attributes together.
Moreover, the insensitive version of MDAV required for 
the method in~\cite{VLDB} to 
fulfill differential privacy produces yet less 
homogeneous clusters, due to the artificial
total order enforced for input records. 

When comparing both differentially private approaches, 
our current proposal offers a significant
improvement in most cases (for RE and JSD), 
but such an improvement depends on the number 
of attributes to be protected.
For $m=1$, we observe the largest differences between both methods, because
the scale parameter of our approach is just $\Delta(A)/(k\times\varepsilon)$
whereas it is $n/k \times (\Delta(A)/(k\times\varepsilon)$ 
for~\cite{VLDB}.
In fact, even for the smallest $\varepsilon$ value ($\varepsilon=0.1$), 
the reduction of information loss is noticeable.
However, as shown also in the previous experiments, for the 
largest $\varepsilon$ value ($\varepsilon=10$),
the distortion caused by individual ranking dominates 
the small amount noise subsequently added to satisfy
differential privacy, which is especially noticeable for the JSD;
for $k>20$ (for RE) or $k>10$ (for JSD) this effect overrides 
the improvement brought by our proposed method.

When the number $m$ of attributes to be protected increases, 
the effective reduction of RE and JSD
achieved by our approach decreases, because
the scale of the required noise increases as
 $\Delta(A)/(k\times(\varepsilon/m))$,
whereas it stays constant for the method in~\cite{VLDB}. 
Indeed, using prior individual-ranking microaggregation requires subsequently
adding noise whose scale factor increases with $m$ in order
to attain $\varepsilon$-differential privacy, whereas
the noise to be added after multivariate microaggregation
does not depend on $m$ (see Section~\ref{ranking} above). When $m$
grows, this subsequent noise addition may override the advantage
of individual ranking with respect to multivariate microaggregation;
in practice, this only happens in pretty extreme cases
(with many attributes, small $\varepsilon$ values and 
large $k$ values), because
multivariate microaggregation also incurs substantial information loss
for large $m$. 
In Figure~\ref{figIRkDP-Census} we can see some such extreme cases
($m=13$, $\varepsilon=10$ and $k>80$ values) in which the 
method in~\cite{VLDB} is able to match or slightly outperform
our method; this is more noticeable for JSD than for RE
because, in the former case, the discretization of the 
continuous attribute values
into bins smooths the effects of the noise.
It is however interesting to observe that, for $\varepsilon=10$, 
our method is even able to outperform the multivariate
microaggregation for $k>20$ (with 8 attributes) and for $k>30$ (with 13 attributes).
This shows the greater room for improvement that individual 
ranking offers over the multivariate microaggregation for $m>1$. 

In fact, by equating the noise scale parameter of~\cite{VLDB}
and our method, we have 
\[ n/k \times (\Delta(A_i)/(k\times\varepsilon) = 
m \times (\Delta(A_i)/(k\times\varepsilon). \]
Hence, both scale parameters coincide for $n/k=m$. Thus,
for $k > n/m$, the scale parameter of~\cite{VLDB} and therefore
the noise added to reach $\varepsilon$-differential privacy are smaller.
For the ``Census'' data set, this means
$k > (1,080/13) \approx 83$.
When the number of attributes to be protected decreases to $m=8$,
we get $k > (1,080/8) = 135$ (which is beyond the $k$ values
represented in the X-axes of the above figures); further, for $m=1$,
we get $k > (1,080/1)=n$, which is outright impossible.
Moreover, the practical reduction of information loss is only noticeable 
for small $\varepsilon$ values (0.1 and 1.0). For $\varepsilon=10.0$, the
high information loss caused by the insensitive MDAV algorithm alone
severely limits the noise reduction gain.

Finally, to evaluate the influence of the prior microaggregation
algorithm, we measured the preservation of
attribute variances as described in Section~\ref{sec:eval1}.
This was done for ``Census'', which is the only
data set with continuous numeric attributes.
The results with 8 attributes are reported in Table~\ref{stats-num}.
In particular, the table shows that,
for medium and large $\varepsilon$ (values 1 and 10),
our method based on individual ranking
significantly improves
on the MDAV-based one in~\cite{VLDB} for all $k$.

\begin{table*}
\caption{\label{stats-num}``Census'': variations of attribute variances between the attributes in the original and masked data sets with the proposed method 
(IR, individual ranking) and the one in~\cite{VLDB} (MV, MDAV multivariate microaggregation) using different values of $k$ and $\varepsilon$.}
\centering
{\renewcommand{\arraystretch}{0.95}
\begin{tabular}{|l|c|cc|cc|cc|}\hline
& & \multicolumn{2}{c|}{$\varepsilon=0.1$} & \multicolumn{2}{c|}{$\varepsilon=1$} & \multicolumn{2}{c|}{$\varepsilon=10$}\\
 & $k$ & IR & MV & IR & MV & IR & MV\\
\hline
$\Delta(\sigma^2(A_{1}))$ & 2 & 25.89  & 25.06  & 22.84  & 25.01 & 9.42 & 24.77  \\
	& 25 		& 21.74  & 24.57  & 8.08 & 20.19 & 0.16 & 3.06 \\
 & 50 	& 20.75  & 23.05  & 2.33 & 9.80 & 0.02 & 0.45 \\
 & 100 	& 15.27  & 17.38  & 0.38 & 0.95 & 0.02 & 0.89 \\
\hline
$\Delta(\sigma^2(A_{2}))$  & 2 			& 9.14  & 8.21  & 7.40 & 8.20 & 3.45 & 8.12 \\
	& 25 		& 8.33  & 8.05  & 2.30 & 6.62 & 0.12 & 1.23 \\
 & 50 	& 7.41  & 7.52  & 1.13 & 3.63 & 0.04 & 0.12 \\
 & 100 	& 5.69  & 5.84  & 0.11 & 0.50 & 0.05 & 0.18 \\
\hline
$\Delta(\sigma^2(A_{3}))$  & 2 			& 24.97  & 24.90  & 11.79 & 22.89 & 0.47 & 10.36 \\
	& 25 		& 24.83  & 24.80  & 12.85 & 22.72 & 0.44 & 8.63 \\
 & 50 	& 24.17  & 24.83  & 9.64 & 22.56 & 0.37 & 8.10 \\
 & 100 	& 23.63  & 24.78  & 9.60 & 22.27 & 0.1 & 6.84 \\
\hline
$\Delta(\sigma^2(A_{4}))$  & 2 			& 10.46  & 9.58  & 8.69 & 9.55 & 3.72 & 9.46 \\
	& 25 		& 9.43  & 9.35  & 2.39 & 7.64 & 0.08 & 1.13 \\
 & 50 	& 8.29  & 8.78  & 1.16 & 3.81 & 0.00 & 0.06 \\
 & 100 	& 5.78  & 6.65  & 0.48 & 0.42 & 0.03 & 0.23 \\
\hline
$\Delta(\sigma^2(A_{5}))$  & 2 			& 16.71  & 15.87  & 14.50 & 15.84 & 6.59 & 15.62 \\
	& 25 		& 14.35  & 15.54  & 5.33 & 12.87 & 0.16 & 2.22 \\
 & 50 	& 13.13  & 14.60  & 2.41 & 6.63 & 0.01 & 0.06 \\
 & 100 	& 9.86  & 11.48  & 0.18 & 0.94 & 0.05 & 0.27 \\
\hline
$\Delta(\sigma^2(A_{6}))$  & 2 			& 22.07  & 21.22  & 19.30 & 21.21 & 8.20 & 21.04 \\
	& 25 		& 19.50  & 20.73  & 6.36 & 17.12 & 0.19 & 2.37 \\
 & 50 	& 15.61  & 19.44  & 3.63 & 8.05 & 0.03 & 0.15 \\
 & 100 	& 10.20  & 14.81  & 0.45 & 0.85 & 0.04 & 0.44 \\
\hline
$\Delta(\sigma(^2A_{7}))$  & 2 			& 8.61  & 7.69  & 6.91 & 7.69 & 3.12 & 7.57 \\
	& 25 		& 7.63  & 7.54  & 2.97 & 6.14 & 0.03 & 1.12 \\
 & 50 	& 6.57  & 7.05  & 0.95 & 3.25 & 0.02 & 0.06 \\
 & 100 	& 4.95  & 5.44  & 0.30 & 0.41 & 0.01 & 0.20 \\
\hline
$\Delta(\sigma^2(A_{8}))$  & 2 			& 70.22  & 69.71  & 64.01 & 69.66 & 24.06 & 68.89 \\
	& 25 		& 63.42 & 68.21  & 14.90 & 55.85 & 0.05 & 5.93 \\
 & 50 	& 49.28  & 64.05  & 3.66 & 25.56 & 0.12 & 0.25 \\
 & 100 	& 41.36  & 47.38  & 0.78 & 2.05 & 0.21 & 0.81 \\
\hline
\end{tabular}}
\end{table*}

We can see that the variations of the \textit{attribute variances} are 
always greater (for all attributes and all $k$ and $\varepsilon$ values) 
for the method in~\cite{VLDB} than for the one proposed in this paper.
Differences between the two methods are greater for larger $\varepsilon$ values
(\emph{i.e.}, 1 and 10) and larger $k$ values (\emph{i.e.} 25 and above);
in these cases, the amount of noise that needs to be added
is smaller and, thus, the influence of the prior microaggregation is
more noticeable.
The differences between both methods illustrate 
how individual-ranking microaggregation 
does a better job at preserving the internal structure 
(and, thus the statistical properties) of the attributes, since
these are aggregated independently. In contrast, multivariate
microaggregation is more constrained 
because all the attributes of each 
record are considered at once; this suppresses
more variance and hence incurs higher information loss. 
In any case, the variations of the attribute variances tend to decrease as 
$k$ grows, which suggests that the prior microaggregation helps decrease the large
variance introduced by the noise added to satisfy differential privacy.
However, there are some cases (\emph{i.e.} attributes 2, 4, 5 and 8) 
in which, for large $\varepsilon$ and $k$ values (\emph{i.e.} 10 and 100, respectively),
variations increase; this shows how, for relaxed values of $\varepsilon$,
the distortion introduced by the coarser microaggregation dominates
the reduction of noise.

\section{Conclusions}
\label{conclusions}

In~\cite{VLDB} a method was presented that combines $k$-anonymity  
and $\varepsilon$-differential privacy to reap the 
best of both models: namely, on the one side the 
reasonably low information loss incurred by $k$-anonymity
and its lack of assumptions on data uses (which do not limit
the kind of analyses that can be performed),
and on the other side the robust privacy guarantees 
offered by $\varepsilon$-differential
privacy. In this paper, we have offered an alternative method that, by
relying 
individual ranking
microaggregation,
is able to effectively
reduce even more the scale parameter of noise in most scenarios,
which is of utmost importance for data analysis.
Such noise reduction has been discussed theoretically and it 
has been illustrated 
empirically for two reference data sets, by focusing on the 
error introduced in the attribute values and the preservation 
of attribute distributions. 

The method proposed here is also easier to implement than the 
one in~\cite{VLDB}, because the individual ranking
algorithm only relies on the natural order of individual attributes. 
Moreover,
its computational cost is $O(n\times m)$, 
whereas insensitive multivariate microaggregation takes $O(n\times n)$.
Since usually $n \gg m$, the current method is more scalable
as the number of records in a data set grows. 
Moreover, prior individual-ranking microaggregation incurs
less information loss than the prior multivariate 
microaggregation used in~\cite{VLDB}.
Finally, the proposed method is especially indicated 
when only a subset of attributes needs to be protected
(\emph{e.g.} the confidential attributes).

We leave as future work the exploration of other types of noise. 
For instance,
using a noise calibrated to the smooth sensitivity of the data 
would seem an interesting
improvement. The main reason is that 
it would reduce the dependency of the amount of required noise on the
size of the attribute domain.

\section*{Acknowledgments and disclaimer}

This work was partly supported by the European Commission 
(through project H2020 ``CLARUS''), 
by the Spanish Government (through projects ``ICWT'' TIN2012-32757,
``CO-PRIVACY'' TIN2011-27076-C03-01 and ``SmartGlacis'') 
and by the Government of Catalonia (under grant 2014 SGR 537). 
Josep Domingo-Ferrer is partially supported as an ICREA-Acad\`emia 
researcher by the Government of Catalonia. 
The opinions expressed in this paper 
are the authors' own and do not necessarily reflect the views of 
UNESCO.

\end{document}